%% file: article.tex
\documentclass[9pt,sigconf,screen,authorversion,nonacm]{acmart}

\usepackage{paralist}
\usepackage{makecell}
\usepackage{adjustbox}
\usepackage{booktabs}
\usepackage{enumitem}
\usepackage{fancyhdr}

\AtBeginDocument{%
  \providecommand\BibTeX{{%
    \normalfont B\kern-0.5em{\scshape i\kern-0.25em b}\kern-0.8em\TeX}}}

\setcopyright{acmcopyright}

\acmYear{2023}\copyrightyear{2023}
\acmConference[WoSC '23]{9th International Workshop on Serverless Computing}{December 11--15, 2023}{Bologna, Italy}
\acmBooktitle{9th International Workshop on Serverless Computing (WoSC '23), December 11--15, 2023, Bologna, Italy}
\acmPrice{15.00}
\acmDOI{10.1145/3631295.3631403}
\acmISBN{979-8-4007-0455-0/23/12}

\graphicspath{{figures/}}

\newcommand{\code}[1]{{\small \texttt{#1}}}

\AtBeginDocument{%
    \addtolength{\footskip}{2.0\baselineskip}%
    \fancyfoot[L]{\textit{CC-BY 4.0 --- This is the author's pre-print version. The published article can be found at \cite{10.1145/3631295.3631403} \url{https://dl.acm.org/doi/10.1145/3631295.3631403}.}}%
}

\begin{document}

\title{Scaling a Variant Calling Genomics Pipeline with FaaS}

\author{Aitor Arjona}
\affiliation{%
  \institution{Universitat Rovira i Virgili}
  \city{Tarragona}
  \country{Spain}
}
\email{aitor.arjona@urv.cat}

\author{Arnau Gabriel-Atienza}
\affiliation{%
  \institution{Universitat Rovira i Virgili}
  \city{Tarragona}
  \country{Spain}
}
\email{arnau.gabriel@urv.cat}

\author{Sara Lanuza-Orna}
\affiliation{%
  \institution{Universitat Rovira i Virgili}
  \city{Tarragona}
  \country{Spain}
}
\email{sara.lanuza@urv.cat}

\author{Xavier Roca-Canals}
\affiliation{%
  \institution{Universitat Rovira i Virgili}
  \city{Tarragona}
  \country{Spain}
}
\email{xavier.rocai@urv.cat}

\author{Ayman Bourramouss}
\affiliation{%
  \institution{Universitat Rovira i Virgili}
  \city{Tarragona}
  \country{Spain}
}
\email{ayman.bourramouss@urv.cat}

\author{Tyler K. Chafin}
\affiliation{%
  \institution{Biomathematics and Statistics Scotland}
  \city{Edinburgh}
  \country{United Kingdom}
}
\email{tyler.chafin@bioss.ac.uk}

\author{Lucio Marcello}
\affiliation{%
  \institution{Biomathematics and Statistics Scotland}
  \city{Edinburgh}
  \country{United Kingdom}
}
\email{lucio.marcello@bioss.ac.uk}

\author{Paolo Ribeca}
\affiliation{%
  \institution{Biomathematics and Statistics Scotland}
  \city{Edinburgh}
  \country{United Kingdom}
}
\email{paolo.ribeca@bioss.ac.uk}

\author{Pedro Garc\'ia-L\'opez}
\affiliation{%
  \institution{Universitat Rovira i Virgili}
  \city{Tarragona}
  \country{Spain}
}
\email{pedro.garcia@urv.cat}

\renewcommand{\shortauthors}{Arjona, et al.}

\begin{abstract}
  With the escalating complexity and volume of genomic data, the capacity of biology institutions' HPC faces limitations. While the Cloud presents a viable solution for short-term elasticity, its intricacies pose challenges for bioinformatics users. Alternatively, serverless computing allows for workload scalability with minimal developer burden. However, porting a scientific application to serverless is not a straightforward process. In this article, we present a Variant Calling genomics pipeline migrated from single-node HPC to a serverless architecture. We describe the inherent challenges of this approach and the engineering efforts required to achieve scalability. We contribute by open-sourcing the pipeline for future systems research and as a scalable user-friendly tool for the bioinformatics community.
\end{abstract}




\maketitle

\input{intro}

\input{pipeline}

\input{inputdata}

\input{synchronism}

\input{parallelism}

\input{reduce}

\input{evaluation}

\input{relatedwork}

\input{conclusion}

\begin{acks}
  This work has been partially funded by the EU Horizon programme under grant agreements No. 101086248, No. 101092644, No. 101092646, No. 101093110 and by the Spanish Ministry of Science and Innovation and State Research Agency (Agencia Estatal de Investigaci\'on) under grant agreement PID2019-106774RB-C22. Aitor Arjona is a URV Mart\'i Franqu\`es grant fellow.
\end{acks}

\bibliographystyle{ACM-Reference-Format}
\bibliography{article}

\end{document}

%% file: intro.tex
\section{Introduction}

Genome sequence analysis is a compute- and data-intensive task. Current sequencing techniques generate an overwhelming volume of data, with repositories accumulating several petabytes each year. The European Nucleotide Archive alone surpassed 40 petabytes of data in 2021~\cite{ena-2022}, and this trend is expected to continue in the future~\cite{ebi-ena}. Consequently, institutions struggle to meet the ever-growing demands for genomics workloads. Although High Performance Computing (HPC) installations are commonplace in biology departments, they may fall short in providing sufficient capacity to handle peak demands, reanalyze extensive public datasets generated by large consortia, or minimize the run time needed for specific analyses.

Faced with this situation, many institutions often turn to public Clouds in seek of compute and storage resource elasticity. The Cloud allows to meet exceptional capacity needs on demand, while avoiding the burden of acquiring and maintaining physical infrastructure on a permanent basis. Although there are genomics-centered distributed computing frameworks available (e.g., Next-flow~\cite{di2017nextflow} or GATK for Apache Spark), they present numerous challenges for bio-informatics users. Evaluating, configuring, deploying and scaling the necessary resources and services for each application becomes a daunting task due to the multitude of heterogeneous Cloud providers and the complexities of their services.


\begin{figure*}[htb!]
    \centering
    \includegraphics[width=0.975\linewidth]{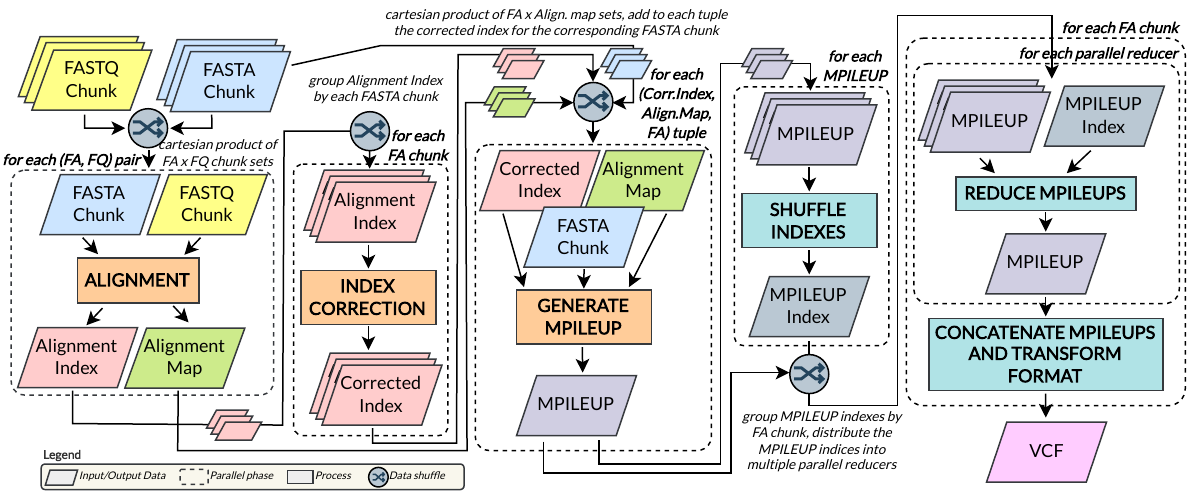}
    \caption{Big picture of the serverless Variant Calling workflow.}
    \label{fig:pipeline-big-picture}
\end{figure*}

\textit{Function-as-a-Service} (FaaS) offers a compelling alternative for highly parallel data-intensive applications~\cite{jonas2017occupy, pu2019shuffling, shankar2018numpywren}, with minimal configuration burden with the ability to instantly scale up and \textit{down to zero}, making it an ideal choice for less experienced Cloud users to deploy workloads. Recent advancements in FaaS services, such as Amazon Lambda's increased resource allocation\footnote{{\url{https://aws.amazon.com/about-aws/whats-new/2020/12/aws-lambda-supports-10gb-memory-6-vcpu-cores-lambda-functions/}}}, open up new possibilities for tackling demanding workloads with serverless. However, adopting the serverless paradigm, despite its cost-effectiveness and scalability benefits, introduces its own set of challenges~\cite{hellerstein2018serverless}. While some efforts have been made to ease this transition~\cite{spillner2018faaster}, it remains a complex task, often necessitating substantial application re-architecting.

Here we present a case study involving the porting of a genomics Variant Calling workflow from HPC to a serverless architecture. Our focus lies on demonstrating the key challenges that arise from this approach, particularly those concerning concurrency, input data partitioning, and stateful data dependencies. The objective of our work is to harness the massive and instant parallelism of FaaS~\cite{barcelona2021benchmarking} to achieve high scalability in a genomic variant calling pipeline, all while ensuring a streamlined experience with minimal burden for the end user. Despite the large data movements involved, by decomposing the workflow into fine-grained tasks and leveraging parallelism in multi-CPU functions, we can ultimately achieve high performance and scalability in a cost-effective way.

Our primary contribution lies in the insights gained from the engineering efforts undertaken during the migration. In summary, serverless enables to massively scale scientific workloads, but adapting them from HPC requires thorough adjustment and complex workflow restructuring. We've open-sourced this work, on GitHub\footnote{\url{https://github.com/CLOUDLAB-URV/serverless-genomics-variant-calling}}, both for the evaluation of serverless research systems and as an open-source tool for scalable bioinformatics variant calling in the Cloud.

%% file: pipeline.tex
\section{Serverless Variant Calling pipeline}\label{sec:pipeline}

Variant calling is a key analysis step in genomics. It typically involves ``alignment'' (i.e., a string similarity search) of sequencing reads stored as FASTQ files to a reference genome stored as a FASTA file. Alignment mismatches are filtered and form the basis for establishing (``calling'') any mutations (``variants'') in the samples analyzed. In general, alignment is a stateful computation requiring the storage into memory of large data structures (``indices'') allowing for fast similarity searches in the reference genome, which in this scenario might be as large as the human genome, or $\approx 3$ billion DNA characters, resulting in indices as large as 5-50 GB depending on the alignment algorithm chosen.

In order to scale, we must distribute the workload, but requiring each serverless function to store a full index for the reference genome is unrealistic due to the memory constraints. However, one can split the FASTA reference genome into smaller chunks and align reads to all of them, at the price of making the computation less efficient and having to introduce a reduce step to collect and score all alignments for each read later on. As each sequencing read is aligned to the genome independently, one can also split the original FASTQ file into smaller chunks in order to trivially increase parallelism. The resulting architecture is shown in Figure~\ref{fig:pipeline-big-picture}.

The original HPC implementation involved bash scripts and CLI tools that utilized the local file system and pipe commands to pass data between processes. To transition this pipeline into a serverless architecture, we employed Python code to wrap the genomic CLI tools and introduce new logic for data partitioning and load distribution across fine-grained serverless tasks. The implementation required 5588 LOC on top of the genomic tools of combined Python, AWK and Bash scripts. This serverless architecture follows PyWren's~\cite{jonas2017occupy} serverless map-reduce model, featuring multiple parallel stages with object storage for input and intermediate data passing. We leveraged the Lithops framework~\cite{sampe2021lithops} for deploying functions on AWS Lambda and used AWS S3 for storage. More in detail, there are three main pipeline phases:

\textbf{\textit{Pre-processing phase:}} This phase prepares the data so that it can be consumed in a parallel and distributed manner. First, indices are generated to allow random access read to FASTA files (more details in Section~\ref{sec:partitioning}). Secondly, an indexed version of the genome partition is generated by the GEM3 Indexer~\cite{sola2012gem} to speed up the following alignment of the sequencing reads.

\textbf{\textit{Map phase:}} This phase consists of aligning the sampled read to the reference genome. Having partitioned both, this results in a cartesian product between the FASTA and FASTQ partition sets. All FASTA\texttimes~FASTQ chunk combinations are aligned using GEM3 mapper~\cite{sola2012gem}. The indexed FASTA chunk is used as the reference for alignment of a given FASTQ chunk, generating an alignment file in \code{.map} format. Because any given alignment arises from a partial scan of the genome (a given FASTA chunk), we must add an extra step (index correction) to discard suboptimal alignments arising from the mapping of any given read to multiple genome chunks. Alignments are converted to the \code{mpileup} format, which collects information from every read to provide a snapshot of the cumulative alignment results for each chromosome position, providing the number of reads supporting each base call detected across all overlapping alignments.

\textbf{\textit{Reduce phase:}} This phase involves a shuffle of all alignment results for variant calling. To distribute the load, the map output is distributed over multiple parallel reduce functions. Data processing involves merging \code{mpileup} data across sets of functions that share the same reference FASTA chunk, i.e.~those files that share the same chromosome positions. The merged \code{mpileup}s are used as input for the SiNPle variant caller~\cite{sinple}. The partial output from each reduce function is then concatenated to produce the final output file. This can be used for any subsequent nucleotide mutation analysis workflow.

Porting this pipeline to a serverless architecture for scalability poses several technical challenges, including the need for effective and efficient FASTA and FASTQ file partitioning with appropriate partition sizes, dealing with blocking code due to data dependencies, and managing stateful data shuffling. In the following sections, we describe how we addressed these challenges.

%% file: inputdata.tex
\section{Unstructured data partitioning}\label{sec:partitioning}

Effective data partitioning is crucial to fully exploit parallelism on serverless workloads. When decomposing a workload into fine-grained serverless tasks, we must provide the partitioned data with fine granularity to avoid bottlenecks in input data consumption. However, unstructured data formats present challenges when accessed in parallel. As it contains sequences of variable length, accessing an arbitrary portion of a FASTA file presents the challenge of establishing the correct sequence identifiers and relevant offsets.

Splitting a large FASTA file into many smaller files is not ideal as it implies reading, partitioning, and writing back to storage the entire file, which can become costly and complex for large files. In addition, a fixed partition size is not ideal, as it will depend on the requirements of each workload. As an alternative, we propose FASTA indexing to enable data partitioning directly from Cloud Object Storage. Our approach involves performing pre-processing to generate indexes for each FASTA file required. This pre-processing is read-only, meaning that data is only read from storage, and only the generated index is written to storage, leaving the original file intact. These indexes are reusable, meaning that they can be used for different partition sizes for different workflows. The generated indexes only consume a minimal portion of storage, significantly reducing the cost and complexity of data partitioning.

We implement an inverse index for FASTA files that allows to lookup contextual genome information of any arbitrary partition, regardless of its size or offset. The index is based on Samtools \code{faidx}~\footnote{\url{https://www.htslib.org/doc/samtools-faidx.html}}. The read-only indexing process leverages parallel serverless functions to pre-process large FASTA files which are kept as-is in object storage. When partitioning a FASTA file, we leverage byte-range GET requests and look up the inverse index to annotate each partition with the required genome metadata. Thanks to this approach, we avoid downloading the entire sequence and storing duplicated partitioned data on object storage.

%% file: synchronism.tex
\section{Data dependencies and synchronous code}

In HPC, a fixed number of parallel processes is commonly maintained throughout the workflow (e.g.~rank size in MPI). This, however, clashes with many FaaS services where only concurrency is guaranteed instead of parallelism, potentially causing related functions to not execute simultaneously. This limitation presents challenges when porting stateful, synchronous blocking HPC code to a serverless architecture, mainly due to task synchronization and blocking function calls (e.g.~all gather in MPI). Blocking serverless functions can be risky due to concurrency limits: the function responsible for releasing resources may not be allocated or may fail, potentially resulting in a deadlock. Additionally, FaaS systems lack direct function communication, necessitating a \textit{rendezvous} service for synchronization, like a Redis server on a Virtual Machine, defeating the whole purpose of a \textit{serverless} architecture.

In our variant calling workflow, the Alignment tasks involved blocking calls to wait for the Index Correction tasks to be run in background. An initial implementation with blocking functions, utilizing Redis for rendezvous and temporary data exchange proved impractical due to scalability limitations. Primarily, large data inputs led to an excessive number of alignment tasks, which easily surpassed the concurrency limit of the FaaS service. This prevented the invocation of the index correction functions, resulting in a deadlock. Additionally, relying on a single instance of Redis proved to be a bottleneck.

To address the aforementioned issues, our implementation consists only of asynchronous tasks that can be scheduled independently while treating FaaS as a work queue. Tasks involving synchronization are divided into two steps, ensuring that there are no blocking calls during execution. However, due to the stateless nature of FaaS, intermediate data must be stored in S3 storage for subsequent functions. Consequently, our pipeline is highly scalable. Through object storage for intermediate data, we provide temporal decoupling of tasks. Strict parallelism is no longer required, making the concurrency guarantees of FaaS systems sufficient. On the negative side, this approach results in increased data movements, potentially leading to longer execution times and higher costs.

%% file: parallelism.tex
\section{Intra-function parallelism}

With FaaS providers enabling the configuration of multiple vCPUs per function, new possibilities arise for exploiting this capability by incorporating parallel data processing within each function instance. In our pipeline, the \code{gem-mapper} tool used in the Alignment step allows to leverage multiple threads for parallel processing. Thus, our focus lies in investigating the possible performance gains and challenges presented by this approach.

We can discern the following implications of intra-function parallelism.

\begin{inparaenum}[(i)]
    \item Utilizing \textbf{intra-function parallelism reduces function invocations} while maintaining parallelism, and enhances data processing efficiency for multi-CPU tasks. This, in turn, lowers the total function invocation count, creating greater room for invoking more functions and processing larger datasets while staying within the FaaS platform's concurrency limit.
    \item Intra-function parallelism efficiency is bounded to the \textbf{parallelizable parts of the process} (Amdahl's Law~\cite{rodgers1985amdhal}), which requires a thorough workflow analysis to avoid over-provisioning functions for tasks that won't benefit from multiprocessing.
    \item Assigning more data to each function results in an \textbf{increase of transferred data volume per function}. This can result in higher overheads caused by the high latency of object storage.
\end{inparaenum}

We evaluated different multi-CPU function configurations to assess the performance of intra-function parallelism of the Alignment step. We measured run time and cost across all configurations using a fixed input sample data volume. This data was evenly distributed across all functions, each configured with the same memory and CPU.

\begin{figure}
    \centering
    \includegraphics[width=\linewidth]{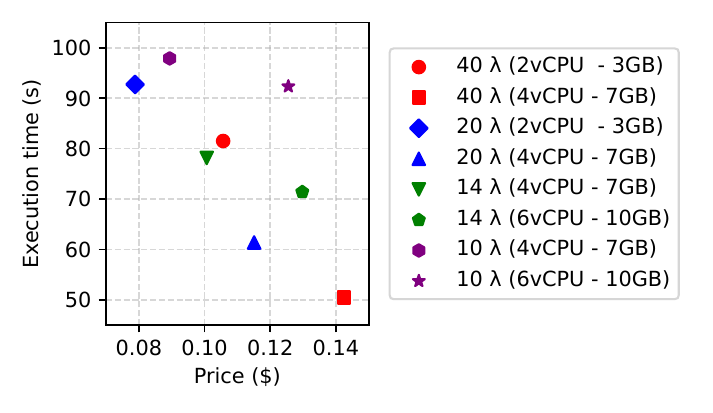}
    \caption{Multiprocessing performance analysis on AWS Lambda using different CPU-Memory configurations}
    \label{figure:big-lambda-experiments}
\end{figure}

Figure~\ref{figure:big-lambda-experiments} presents the results of the experiment regarding the cost-performance relationship. Configurations with a higher number of invocations and greater vCPU resources exhibit reduced execution times, albeit at the cost of higher expenses. However, these configurations require a larger number of function invocations. For large input datasets, this may surpass the FaaS platform's maximum concurrency limit (e.g., 1000 in AWS Lambda), potentially hampering speedup and overall performance gains. Conversely, configurations featuring more vCPUs and larger data volumes per function tend to be more cost-effective but slower.

\begin{figure}
    \centering
    \includegraphics[width=\linewidth]{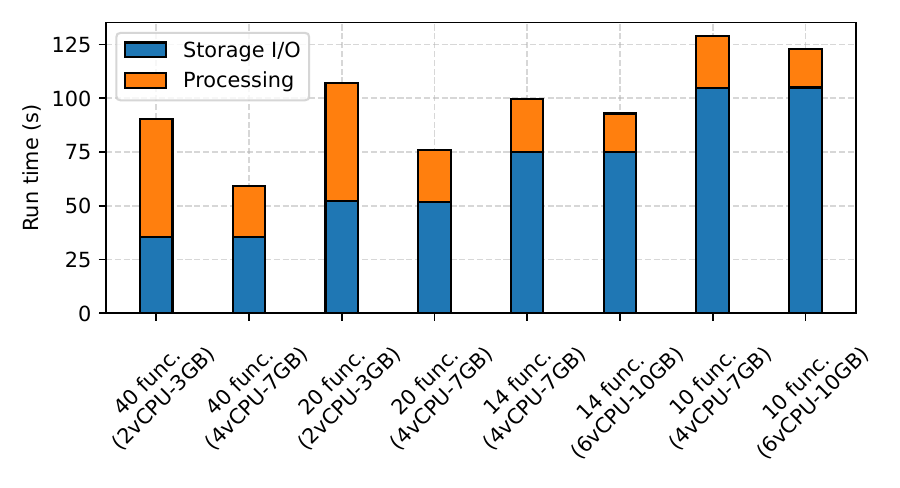}
    \caption{Average data transfer times for a single function.}\label{figure:big-lambda-data-transfers}
\end{figure}

For a more comprehensive analysis of this behavior, we quantified the average run time of storage access and parallel processing across all configurations. The results are displayed in Figure~\ref{figure:big-lambda-data-transfers}. As the number of parallel functions decreases, data download time increases due to a larger data partition size for each function, consuming more execution time. On the other hand, increasing vCPUs per function leads to decreased execution time, indicating efficient intra-function parallelism utilization of the genomic process.

Efficiently leveraging multi-CPU functions is linked to the function-executing process's parallelism capacity, which varies for different workloads and cannot be universally applied. In this specific case, a configuration with 4 vCPUs and 7 GB of memory strikes a good balance between performance and cost.

%% file: reduce.tex
\section{Stateful data movements}

The reduce phase (Variant Calling) of our pipeline involves the merging of \code{mpileup} files pertaining to the same FASTA chunk, then proceesed by SiNPle~\cite{sinple} to provide variant calls, which are then concatenated. To ensure scalability and due to function memory constraints, we need to distribute the computation across different functions. Thus, an initial shuffling phase is needed to determine data partitions for each function.

The complexity of the shuffling lies in the nature of the \code{mpileup} format, which is composed of reads from the different genome positions found in FASTQ that have been compared to the reference FASTA. This process necessitates grouping genome positions (indexes) across multiple \code{mpileup} files pertaining to the same FASTA reference chunk. However, not all indexes may exist in every file, and their quantity can vary. The shuffling algorithm establishes a set of ranges of indices that will form the partitions. The size of the partitions is defined by the memory allocated for each serverless function. This size can be variable depending on the number of functions to invoke capable of processing all the intermediate \code{mpileup} sets. To determine the partition size, we sampled lines to calculate the minimum and maximum size, so we can calculate an approximation of the size of the ranges in bytes to fit into the function's memory. Small index ranges are grouped together in a function, in order to keep the number of invocations low and avoid reaching the concurrency limit. Once the indexes have been distributed, the respective reduce functions will be invoked. Each function will fetch the ranges from the \code{mpileup} file set assigned to it and then apply the processing using SiNPle.

We leverage S3 SELECT for the shuffling process, as it allows to perform SQL queries over the contents of the structured text format of the \code{mpileup} files. First, we query each \code{mpileup} set to extract the column of indexes of each one of the files to be able to compute lines and to establish in a simple way the ranges of indexes for each one of the partitions. Second, each reduce function queries the file to extract directly the lines that are within the range of indexes established in the previous phase. This service helps us avoid the need for additional processing, which would introduce computational complexity and time overhead. Specifically, this simplifies the map phase when pre-processing indexes and during the reduce phase when creating partitions. On the down side, the implementation we propose is not transparent to multiple cloud providers making it totally dependent on Amazon Web Services.

%% file: evaluation.tex
\section{Evaluation}

After a comprehensive review of the serverless Variant Calling pipeline architecture and its implementation details, we will now evaluate two key aspects:
\begin{inparaenum}[(i)]
    \item \textbf{Performance evaluation} where we will compare its execution time to the single-node HPC version, and
    \item \textbf{Scalability evaluation} where we will use a larger dataset to assess the architecture's scalability.
\end{inparaenum}

\begin{figure}
    \centering
    \includegraphics[width=\linewidth]{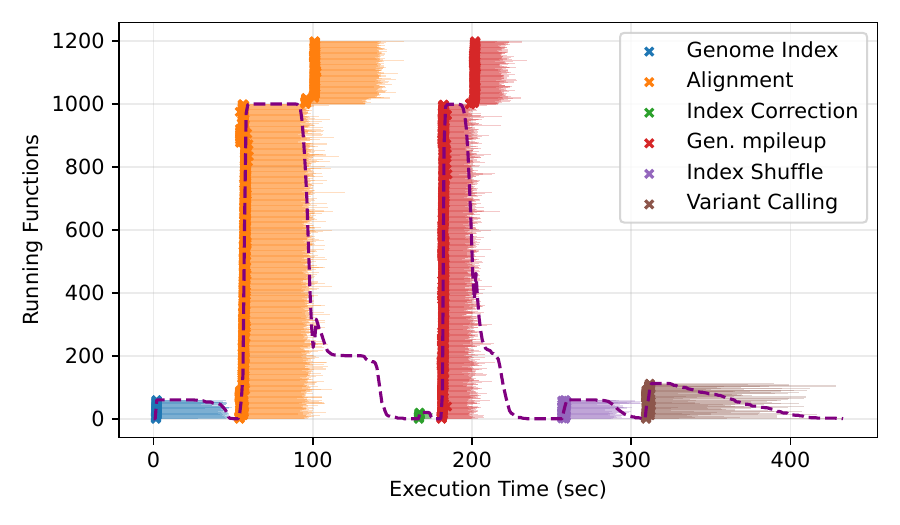}
    \caption{Tasks run time and concurrency.}\label{fig:hg19-histogram}
\end{figure}

\noindent \emph{\textbf{Experiments setup:}} The HPC configuration consists of a Dell PowerEdge R440 node equipped with two Intel Xeon Gold 5118 CPUs @ 2.30GHz, with a total of 24 cores (48 threads) and 32GB of memory. For the serverless AWS Lambda execution, we utilized a Docker function runtime configured with $8192$ MB of memory (which provides 4 vCPUs) and an ephemeral disk size of $10240$ MB. We used Amazon S3 for the storage of the input and intermediate data, with a bucket located in the same region (\code{us-east-1}) as the serverless functions.

\subsection{Performance}

As a first benchmark, we used the Trypanosome genome\footnote{\url{https://tritrypdb.org/tritrypdb/app/downloads/Current_Release/TbruceiTREU927/fasta/data/}} and sequence reads \code{SRR6052133}\footnote{\url{https://trace.ncbi.nlm.nih.gov/Traces/?view=run_browser&acc=SRR6052133&display=download}} from NCBI.


\begin{table}[htb!]
    \caption{Comparison of execution times between HPC and Serverless for the Variant Calling stages.}\label{tab:performance-results}
    \begin{tabular}{@{} m{3cm} m{2.25cm} m{2.25cm} @{}}
        \toprule
        \textbf{\emph{Stage}} & \emph{HPC}     & \emph{Serverless} \\ \midrule
        Genome Indexing       & 0 min 14.20 s  & 0 min 9.81 s      \\
        Alignment             & 0 min 14.20 s  & 0 min 48.10 s     \\
        Index correction      & -              & 0 min 7.63 s      \\
        Generate mpileup      & 51 min 15.79 s & 1 min 6.55 s      \\
        Index shuffle         & -              & 0 min 10.73 s     \\
        Variant Calling       & 54 min 5.04s s & 0 min 27.82 s     \\ \midrule
        \textbf{Total}        & 106 min 8.21 s & 2 min 50.64 s     \\ \bottomrule\hline
    \end{tabular}
\end{table}

Table~\ref{tab:performance-results} presents execution times for the stages detailed in Section~\ref{sec:pipeline}. The HPC version efficiently leverages multithreading for genome indexing and alignment processes but experiences performance bottlenecks in the \code{mpileup} and variant calling phases as they predominantly consist of sequential code. Conversely, the serverless setup optimally distributes workloads, resulting in a significant reduction in overall execution time even for sequential code.

\subsection{Scalability}\label{sec:validation-scalability}

Next, we wanted to assess the scalability of our architecture by using a larger input dataset such as the \code{hg19}\footnote{\url{http://hgdownload.cse.ucsc.edu/goldenpath/hg19/bigZips/}} FASTA reference genome with $20$ partitions and the \code{SRR15068323} FASTQ read set with $60$ partitions from the NCBI archive~\footnote{\url{https://trace.ncbi.nlm.nih.gov/Traces/?view=run_browser&acc=SRR15068323&display=download}}. This represents a realistic benchmark of variant calling in the field of human genomics. The files have an uncompressed size of 3.0 GB and 5.4 GB, respectively.

Figure~\ref{fig:hg19-histogram} illustrates the pipeline's tasks execution time and concurrency over time, using 1200 tasks for the alignment stage and 112 for the reduce stage. Thanks to the asynchronous code redesign, we can comfortably launch more tasks than the available default concurrency of 1000 functions for Lambda. In this case, we launch a total of 1200 parallel functions distributed in two flights of 1000 and 200 invocations respectively. This implies that we can scale and launch even more functions if necessary for larger datasets, regardless of the concurrency limit of the FaaS platform. Furthermore, the FaaS platform adapts to the highly elastic parallelism throughout the pipeline for optimal cost-effectiveness, leaving no idle resources under-utilized. The serverless architecture allowed us to scale this workload, and achieve an execution time of 7 minutes and 16.57 seconds. For reference, the same workload ran for $\approx 14$ hours on the HPC setup. The total execution cost was 10.68 USD for AWS Lambda GB-sec billing, with an additional 0.0644 USD for 32.2GB data scanned in S3 SELECT.

\section{Insights}

The main insights we gained from migrating this variant-calling bioinformatics task to a serverless architecture are:

\begin{enumerate}[wide]
    \item Optimizing performance-cost with \textbf{multi-CPU functions} requires \textbf{balancing intra- and inter-function parallelism}. Achieving a compromise is a non-trivial task, since allocating more CPUs per function is beneficial for tasks that leverage multiprocessing, but at the same time it also significantly increases execution costs.

    \item Effective \textbf{data partitioning} is vital to exploit the massive parallelism offered by FaaS. Data indexing can help in reducing the costs of partitioning, since it prevents writing duplicated partitioned data back to storage. However, a partition size still has to be explicitly configured, which adds further complexity.

    \item An \textbf{asynchronous and non-blocking architecture} is strictly necessary to guarantee scalability and to avoid the concurrency limits of serverless platforms. This requires, however, a deep understanding of the pipeline, and may result in a complex code re-architecting.

    \item Although \textbf{object storage} offers scalability for intermediate workflow data storage, it suffers from \textbf{slow performance and elevated costs}. Its affordable storage cost per GB is completely outweighed by the very expensive blocking I/O from Lambda runtime billing. Specifically, in the experiment detailed in Section~\ref{sec:validation-scalability}, the accumulated I/O time accounts for 18\% of the total runtime during the alignment stage and 34\% during the reduce stage, considering the blocking S3 SELECT calls. Despite object storage being the \textit{state-of-practice} for serverless workloads, there still remains a need for a high-performance, scalable storage service in the Cloud for short-term data storage.

    \item \textbf{Serverless data processing services for object storage}, like S3 SELECT, provide a nice synergy with serverless FaaS workloads. Yet, S3 SELECT is limited to a closed set of SQL expressions over semi-structured files (CSV, JSON or Parquet). We expect Cloud providers to introduce new services similar to S3 Select, but aimed to facilitate scientific data management in serverless architectures.
\end{enumerate}

%% file: relatedwork.tex
\section{Related Work}

The complexity of genomic workloads has led to the exploration of using serverless computing in several works. In~\cite{aji2021evaluation}, the authors use SWEEP to execute a serverless variant calling workflow. SWEEP defines workflows as a directed acyclic graph (DAG) of stateless tasks and manages the deployment of these tasks in serverless backends (functions or containers). Their approach focuses on parallelizing and distributing workflows based on the number of samples --- in contrast, our approach is centered on designing a distributed algorithm to parallelize and scale the analysis of individual samples, ensuring scalability for large sample sizes.

Other works also explored serverless computing for various genomic algorithms, like alignment~\cite{niu2019leveraging}, SNP genotyping data mining~\cite{cepeda2019challenges} or RNA visualization~\cite{lee2019dnavisualization}. These works parallelize the algorithms, but they consist of a single parallel serverless stage, using object storage for input and output data. In the case of~\cite{hong2020accessible} they employed serverless for genome alignment but used virtual machines for stateful computations. In our approach, we implement a fully serverless architecture with different parallel processing stages using object storage for data shuffling and stateful data movements.

%% file: conclusion.tex
\section{Conclusion}

In this article, we have reviewed the implications of moving a single-node HPC workload to a serverless architecture. Although we succeeded in the goal of scalability, it required daunting engineering work to wrap the original code and make it suitable for FaaS. Limitations such as the stateless nature of functions or concurrency and resource constraints proved to be challenging. As a result, we offer the pipeline in open source for future optimization works and as a highly scalable tool for genomic variant calling.

%% file: article.bbl

\begin{thebibliography}{19}


\ifx \showCODEN    \undefined \def \showCODEN     #1{\unskip}     \fi
\ifx \showDOI      \undefined \def \showDOI       #1{#1}\fi
\ifx \showISBNx    \undefined \def \showISBNx     #1{\unskip}     \fi
\ifx \showISBNxiii \undefined \def \showISBNxiii  #1{\unskip}     \fi
\ifx \showISSN     \undefined \def \showISSN      #1{\unskip}     \fi
\ifx \showLCCN     \undefined \def \showLCCN      #1{\unskip}     \fi
\ifx \shownote     \undefined \def \shownote      #1{#1}          \fi
\ifx \showarticletitle \undefined \def \showarticletitle #1{#1}   \fi
\ifx \showURL      \undefined \def \showURL       {\relax}        \fi
\providecommand\bibfield[2]{#2}
\providecommand\bibinfo[2]{#2}
\providecommand\natexlab[1]{#1}
\providecommand\showeprint[2][]{arXiv:#2}

\bibitem[Archive(2022)]%
        {ebi-ena}
\bibfield{author}{\bibinfo{person}{European~Nucleotide Archive}.}
  \bibinfo{year}{2022}\natexlab{}.
\newblock \bibinfo{title}{Statistics}.
\newblock
  \bibinfo{howpublished}{\url{https://www.ebi.ac.uk/ena/browser/about/statistics}}.
\newblock


\bibitem[Arjona et~al\mbox{.}(2023)]%
        {10.1145/3631295.3631403}
\bibfield{author}{\bibinfo{person}{Aitor Arjona}, \bibinfo{person}{Arnau
  Gabriel-Atienza}, \bibinfo{person}{Sara Lanuza-Orna}, \bibinfo{person}{Xavier
  Roca-Canals}, \bibinfo{person}{Ayman Bourramouss}, \bibinfo{person}{Tyler~K.
  Chafin}, \bibinfo{person}{Lucio Marcello}, \bibinfo{person}{Paolo Ribeca},
  {and} \bibinfo{person}{Pedro Garc\'{\i}a-L\'{o}pez}.}
  \bibinfo{year}{2023}\natexlab{}.
\newblock \showarticletitle{Scaling a Variant Calling Genomics Pipeline with
  FaaS}. In \bibinfo{booktitle}{\emph{Proceedings of the 9th International
  Workshop on Serverless Computing}} (Bologna, Italy)
  \emph{(\bibinfo{series}{WoSC '23})}. \bibinfo{publisher}{Association for
  Computing Machinery}, \bibinfo{address}{New York, NY, USA},
  \bibinfo{pages}{59–64}.
\newblock
\showISBNx{9798400704550}
\urldef\tempurl%
\url{https://doi.org/10.1145/3631295.3631403}
\showDOI{\tempurl}


\bibitem[Barcelona-Pons and García-López(2021)]%
        {barcelona2021benchmarking}
\bibfield{author}{\bibinfo{person}{Daniel Barcelona-Pons} {and}
  \bibinfo{person}{Pedro García-López}.} \bibinfo{year}{2021}\natexlab{}.
\newblock \showarticletitle{Benchmarking parallelism in FaaS platforms}.
\newblock \bibinfo{journal}{\emph{Future Generation Computer Systems}}
  \bibinfo{volume}{124} (\bibinfo{year}{2021}), \bibinfo{pages}{268--284}.
\newblock
\showISSN{0167-739X}
\urldef\tempurl%
\url{https://doi.org/10.1016/j.future.2021.06.005}
\showDOI{\tempurl}


\bibitem[Burgin et~al\mbox{.}(2023)]%
        {ena-2022}
\bibfield{author}{\bibinfo{person}{J. Burgin}, \bibinfo{person}{A. Ahamed},
  \bibinfo{person}{C. Cummins}, \bibinfo{person}{R. Devraj},
  \bibinfo{person}{K. Gueye}, \bibinfo{person}{D. Gupta}, \bibinfo{person}{V.
  Gupta}, \bibinfo{person}{M. Haseeb}, \bibinfo{person}{M. Ihsan},
  \bibinfo{person}{E. Ivanov}, \bibinfo{person}{S. Jayathilaka},
  \bibinfo{person}{V. Balavenkataraman~Kadhirvelu}, \bibinfo{person}{M. Kumar},
  \bibinfo{person}{A. Lathi}, \bibinfo{person}{R. Leinonen},
  \bibinfo{person}{M. Mansurova}, \bibinfo{person}{J. McKinnon},
  \bibinfo{person}{C. O'Cathail}, \bibinfo{person}{J. rio}, \bibinfo{person}{S.
  Pesant}, \bibinfo{person}{N. Rahman}, \bibinfo{person}{G. Rinck},
  \bibinfo{person}{S. Selvakumar}, \bibinfo{person}{S. Suman},
  \bibinfo{person}{S. Vijayaraja}, \bibinfo{person}{Z. Waheed},
  \bibinfo{person}{P. Woollard}, \bibinfo{person}{D. Yuan}, \bibinfo{person}{A.
  Zyoud}, \bibinfo{person}{T. Burdett}, {and} \bibinfo{person}{G. Cochrane}.}
  \bibinfo{year}{2023}\natexlab{}.
\newblock \showarticletitle{{{T}he {E}uropean {N}ucleotide {A}rchive in 2022}}.
\newblock \bibinfo{journal}{\emph{Nucleic Acids Res}} \bibinfo{volume}{51},
  \bibinfo{number}{D1} (\bibinfo{date}{Jan} \bibinfo{year}{2023}),
  \bibinfo{pages}{D121--D125}.
\newblock


\bibitem[Crespo-Cepeda et~al\mbox{.}(2019)]%
        {cepeda2019challenges}
\bibfield{author}{\bibinfo{person}{Rodrigo Crespo-Cepeda},
  \bibinfo{person}{Giuseppe Agapito}, \bibinfo{person}{Jose Vazquez-Poletti},
  {and} \bibinfo{person}{Mario Cannataro}.} \bibinfo{year}{2019}\natexlab{}.
\newblock \showarticletitle{Challenges and Opportunities of Amazon Serverless
  Lambda Services in Bioinformatics}. \bibinfo{pages}{663--668}.
\newblock
\showISBNx{978-1-4503-6666-3}
\urldef\tempurl%
\url{https://doi.org/10.1145/3307339.3343462}
\showDOI{\tempurl}


\bibitem[Di~Tommaso et~al\mbox{.}(2017)]%
        {di2017nextflow}
\bibfield{author}{\bibinfo{person}{Paolo Di~Tommaso}, \bibinfo{person}{Maria
  Chatzou}, \bibinfo{person}{Evan~W Floden}, \bibinfo{person}{Pablo~Prieto
  Barja}, \bibinfo{person}{Emilio Palumbo}, {and} \bibinfo{person}{Cedric
  Notredame}.} \bibinfo{year}{2017}\natexlab{}.
\newblock \showarticletitle{Nextflow enables reproducible computational
  workflows}.
\newblock \bibinfo{journal}{\emph{Nature biotechnology}} \bibinfo{volume}{35},
  \bibinfo{number}{4} (\bibinfo{year}{2017}), \bibinfo{pages}{316--319}.
\newblock


\bibitem[Ferretti et~al\mbox{.}(2019)]%
        {sinple}
\bibfield{author}{\bibinfo{person}{L. Ferretti}, \bibinfo{person}{C.
  Tennakoon}, \bibinfo{person}{A. Silesian}, \bibinfo{person}{G. Freimanis},
  {and} \bibinfo{person}{P. Ribeca}.} \bibinfo{year}{2019}\natexlab{}.
\newblock \showarticletitle{SiNPle: Fast and Sensitive Variant Calling for Deep
  Sequencing Data}.
\newblock \bibinfo{journal}{\emph{Genes}} (\bibinfo{year}{2019}).
\newblock


\bibitem[Hellerstein et~al\mbox{.}(2018)]%
        {hellerstein2018serverless}
\bibfield{author}{\bibinfo{person}{Joseph~M. Hellerstein},
  \bibinfo{person}{Jose Faleiro}, \bibinfo{person}{Joseph~E. Gonzalez},
  \bibinfo{person}{Johann Schleier-Smith}, \bibinfo{person}{Vikram Sreekanti},
  \bibinfo{person}{Alexey Tumanov}, {and} \bibinfo{person}{Chenggang Wu}.}
  \bibinfo{year}{2018}\natexlab{}.
\newblock \bibinfo{title}{Serverless Computing: One Step Forward, Two Steps
  Back}.
\newblock
\newblock
\showeprint[arxiv]{1812.03651}~[cs.DC]


\bibitem[Hung et~al\mbox{.}(2020)]%
        {hong2020accessible}
\bibfield{author}{\bibinfo{person}{Ling-Hong Hung}, \bibinfo{person}{Xingzhi
  Niu}, \bibinfo{person}{Wes Lloyd}, {and} \bibinfo{person}{Ka~Yee Yeung}.}
  \bibinfo{year}{2020}\natexlab{}.
\newblock \showarticletitle{Accessible and interactive RNA sequencing analysis
  using serverless computing}.
\newblock \bibinfo{journal}{\emph{bioRxiv}} (\bibinfo{year}{2020}).
\newblock
\urldef\tempurl%
\url{https://doi.org/10.1101/576199}
\showDOI{\tempurl}
\showeprint{https://www.biorxiv.org/content/early/2020/10/03/576199.full.pdf}


\bibitem[John et~al\mbox{.}(2021)]%
        {aji2021evaluation}
\bibfield{author}{\bibinfo{person}{Aji John}, \bibinfo{person}{Kathleen
  Muenzen}, {and} \bibinfo{person}{Kristiina Ausmees}.}
  \bibinfo{year}{2021}\natexlab{}.
\newblock \showarticletitle{Evaluation of serverless computing for scalable
  execution of a joint variant calling workflow}.
\newblock \bibinfo{journal}{\emph{PLOS ONE}} \bibinfo{volume}{16},
  \bibinfo{number}{7} (\bibinfo{date}{07} \bibinfo{year}{2021}),
  \bibinfo{pages}{1--12}.
\newblock
\urldef\tempurl%
\url{https://doi.org/10.1371/journal.pone.0254363}
\showDOI{\tempurl}


\bibitem[Jonas et~al\mbox{.}(2017)]%
        {jonas2017occupy}
\bibfield{author}{\bibinfo{person}{Eric Jonas}, \bibinfo{person}{Qifan Pu},
  \bibinfo{person}{Shivaram Venkataraman}, \bibinfo{person}{Ion Stoica}, {and}
  \bibinfo{person}{Benjamin Recht}.} \bibinfo{year}{2017}\natexlab{}.
\newblock \showarticletitle{Occupy the Cloud: Distributed Computing for the
  99\%}. In \bibinfo{booktitle}{\emph{Proceedings of the 2017 Symposium on
  Cloud Computing}} (Santa Clara, California) \emph{(\bibinfo{series}{SoCC
  '17})}. \bibinfo{publisher}{Association for Computing Machinery},
  \bibinfo{address}{New York, NY, USA}, \bibinfo{pages}{445–451}.
\newblock
\showISBNx{9781450350280}
\urldef\tempurl%
\url{https://doi.org/10.1145/3127479.3128601}
\showDOI{\tempurl}


\bibitem[Lee et~al\mbox{.}(2019)]%
        {lee2019dnavisualization}
\bibfield{author}{\bibinfo{person}{Benjamin~D Lee}, \bibinfo{person}{Michael~A
  Timony}, {and} \bibinfo{person}{Pablo Ruiz}.}
  \bibinfo{year}{2019}\natexlab{}.
\newblock \showarticletitle{{DNAvisualization.org}: a serverless web tool for
  {DNA} sequence visualization}.
\newblock \bibinfo{journal}{\emph{Nucleic Acids Res}} \bibinfo{volume}{47},
  \bibinfo{number}{W1} (\bibinfo{date}{July} \bibinfo{year}{2019}),
  \bibinfo{pages}{W20--W25}.
\newblock


\bibitem[Marco-Sola et~al\mbox{.}(2012)]%
        {sola2012gem}
\bibfield{author}{\bibinfo{person}{Santiago Marco-Sola},
  \bibinfo{person}{Michael Sammeth}, \bibinfo{person}{Roderic Guig{\'o}}, {and}
  \bibinfo{person}{Paolo Ribeca}.} \bibinfo{year}{2012}\natexlab{}.
\newblock \showarticletitle{The GEM mapper: fast, accurate and versatile
  alignment by filtration}.
\newblock \bibinfo{journal}{\emph{Nature methods}} \bibinfo{volume}{9},
  \bibinfo{number}{12} (\bibinfo{year}{2012}), \bibinfo{pages}{1185--1188}.
\newblock


\bibitem[Niu et~al\mbox{.}(2019)]%
        {niu2019leveraging}
\bibfield{author}{\bibinfo{person}{Xingzhi Niu}, \bibinfo{person}{Dimitar
  Kumanov}, \bibinfo{person}{Ling-Hong Hung}, \bibinfo{person}{Wesley Lloyd},
  {and} \bibinfo{person}{Ka~Yee Yeung}.} \bibinfo{year}{2019}\natexlab{}.
\newblock \showarticletitle{Leveraging Serverless Computing to Improve
  Performance for Sequence Comparison}. \bibinfo{pages}{683--687}.
\newblock
\showISBNx{978-1-4503-6666-3}
\urldef\tempurl%
\url{https://doi.org/10.1145/3307339.3343465}
\showDOI{\tempurl}


\bibitem[Pu et~al\mbox{.}(2019)]%
        {pu2019shuffling}
\bibfield{author}{\bibinfo{person}{Qifan Pu}, \bibinfo{person}{Shivaram
  Venkataraman}, {and} \bibinfo{person}{Ion Stoica}.}
  \bibinfo{year}{2019}\natexlab{}.
\newblock \showarticletitle{Shuffling, Fast and Slow: Scalable Analytics on
  Serverless Infrastructure}. In \bibinfo{booktitle}{\emph{16th USENIX
  Symposium on Networked Systems Design and Implementation (NSDI 19)}}.
  \bibinfo{publisher}{USENIX Association}, \bibinfo{address}{Boston, MA},
  \bibinfo{pages}{193--206}.
\newblock
\showISBNx{978-1-931971-49-2}
\urldef\tempurl%
\url{https://www.usenix.org/conference/nsdi19/presentation/pu}
\showURL{%
\tempurl}


\bibitem[Rodgers(1985)]%
        {rodgers1985amdhal}
\bibfield{author}{\bibinfo{person}{David~P. Rodgers}.}
  \bibinfo{year}{1985}\natexlab{}.
\newblock \showarticletitle{Improvements in Multiprocessor System Design}.
\newblock \bibinfo{journal}{\emph{SIGARCH Comput. Archit. News}}
  \bibinfo{volume}{13}, \bibinfo{number}{3} (\bibinfo{date}{jun}
  \bibinfo{year}{1985}), \bibinfo{pages}{225–231}.
\newblock
\showISSN{0163-5964}
\urldef\tempurl%
\url{https://doi.org/10.1145/327070.327215}
\showDOI{\tempurl}


\bibitem[Sampe et~al\mbox{.}(2021)]%
        {sampe2021lithops}
\bibfield{author}{\bibinfo{person}{Josep Sampe}, \bibinfo{person}{Pedro
  Garcia-Lopez}, \bibinfo{person}{Marc Sanchez-Artigas}, \bibinfo{person}{Gil
  Vernik}, \bibinfo{person}{Pol Roca-Llaberia}, {and} \bibinfo{person}{Aitor
  Arjona}.} \bibinfo{year}{2021}\natexlab{}.
\newblock \showarticletitle{Toward Multicloud Access Transparency in Serverless
  Computing}.
\newblock \bibinfo{journal}{\emph{IEEE Software}} \bibinfo{volume}{38},
  \bibinfo{number}{1} (\bibinfo{year}{2021}), \bibinfo{pages}{68--74}.
\newblock
\urldef\tempurl%
\url{https://doi.org/10.1109/MS.2020.3029994}
\showDOI{\tempurl}


\bibitem[Shankar et~al\mbox{.}(2018)]%
        {shankar2018numpywren}
\bibfield{author}{\bibinfo{person}{Vaishaal Shankar}, \bibinfo{person}{Karl
  Krauth}, \bibinfo{person}{Qifan Pu}, \bibinfo{person}{Eric Jonas},
  \bibinfo{person}{Shivaram Venkataraman}, \bibinfo{person}{Ion Stoica},
  \bibinfo{person}{Benjamin Recht}, {and} \bibinfo{person}{Jonathan
  Ragan-Kelley}.} \bibinfo{year}{2018}\natexlab{}.
\newblock \bibinfo{title}{numpywren: serverless linear algebra}.
\newblock
\newblock
\showeprint[arxiv]{1810.09679}~[cs.DC]


\bibitem[Spillner et~al\mbox{.}(2018)]%
        {spillner2018faaster}
\bibfield{author}{\bibinfo{person}{Josef Spillner}, \bibinfo{person}{Cristian
  Mateos}, {and} \bibinfo{person}{David~A. Monge}.}
  \bibinfo{year}{2018}\natexlab{}.
\newblock \showarticletitle{FaaSter, Better, Cheaper: The Prospect of
  Serverless Scientific Computing and HPC}. In \bibinfo{booktitle}{\emph{High
  Performance Computing}}, \bibfield{editor}{\bibinfo{person}{Esteban Mocskos}
  {and} \bibinfo{person}{Sergio Nesmachnow}} (Eds.).
  \bibinfo{publisher}{Springer International Publishing},
  \bibinfo{address}{Cham}, \bibinfo{pages}{154--168}.
\newblock
\showISBNx{978-3-319-73353-1}


\end{thebibliography}



\begin{thebibliography}{18}


\ifx \showCODEN    \undefined \def \showCODEN     #1{\unskip}     \fi
\ifx \showDOI      \undefined \def \showDOI       #1{#1}\fi
\ifx \showISBNx    \undefined \def \showISBNx     #1{\unskip}     \fi
\ifx \showISBNxiii \undefined \def \showISBNxiii  #1{\unskip}     \fi
\ifx \showISSN     \undefined \def \showISSN      #1{\unskip}     \fi
\ifx \showLCCN     \undefined \def \showLCCN      #1{\unskip}     \fi
\ifx \shownote     \undefined \def \shownote      #1{#1}          \fi
\ifx \showarticletitle \undefined \def \showarticletitle #1{#1}   \fi
\ifx \showURL      \undefined \def \showURL       {\relax}        \fi
\providecommand\bibfield[2]{#2}
\providecommand\bibinfo[2]{#2}
\providecommand\natexlab[1]{#1}
\providecommand\showeprint[2][]{arXiv:#2}

\bibitem[Archive(2022)]%
        {ebi-ena}
\bibfield{author}{\bibinfo{person}{European~Nucleotide Archive}.}
  \bibinfo{year}{2022}\natexlab{}.
\newblock \bibinfo{title}{Statistics}.
\newblock
  \bibinfo{howpublished}{\url{https://www.ebi.ac.uk/ena/browser/about/statistics}}.
\newblock


\bibitem[Barcelona-Pons and García-López(2021)]%
        {barcelona2021benchmarking}
\bibfield{author}{\bibinfo{person}{Daniel Barcelona-Pons} {and}
  \bibinfo{person}{Pedro García-López}.} \bibinfo{year}{2021}\natexlab{}.
\newblock \showarticletitle{Benchmarking parallelism in FaaS platforms}.
\newblock \bibinfo{journal}{\emph{Future Generation Computer Systems}}
  \bibinfo{volume}{124} (\bibinfo{year}{2021}), \bibinfo{pages}{268--284}.
\newblock
\showISSN{0167-739X}
\urldef\tempurl%
\url{https://doi.org/10.1016/j.future.2021.06.005}
\showDOI{\tempurl}


\bibitem[Burgin et~al\mbox{.}(2023)]%
        {ena-2022}
\bibfield{author}{\bibinfo{person}{J. Burgin}, \bibinfo{person}{A. Ahamed},
  \bibinfo{person}{C. Cummins}, \bibinfo{person}{R. Devraj},
  \bibinfo{person}{K. Gueye}, \bibinfo{person}{D. Gupta}, \bibinfo{person}{V.
  Gupta}, \bibinfo{person}{M. Haseeb}, \bibinfo{person}{M. Ihsan},
  \bibinfo{person}{E. Ivanov}, \bibinfo{person}{S. Jayathilaka},
  \bibinfo{person}{V. Balavenkataraman~Kadhirvelu}, \bibinfo{person}{M. Kumar},
  \bibinfo{person}{A. Lathi}, \bibinfo{person}{R. Leinonen},
  \bibinfo{person}{M. Mansurova}, \bibinfo{person}{J. McKinnon},
  \bibinfo{person}{C. O'Cathail}, \bibinfo{person}{J. rio}, \bibinfo{person}{S.
  Pesant}, \bibinfo{person}{N. Rahman}, \bibinfo{person}{G. Rinck},
  \bibinfo{person}{S. Selvakumar}, \bibinfo{person}{S. Suman},
  \bibinfo{person}{S. Vijayaraja}, \bibinfo{person}{Z. Waheed},
  \bibinfo{person}{P. Woollard}, \bibinfo{person}{D. Yuan}, \bibinfo{person}{A.
  Zyoud}, \bibinfo{person}{T. Burdett}, {and} \bibinfo{person}{G. Cochrane}.}
  \bibinfo{year}{2023}\natexlab{}.
\newblock \showarticletitle{{{T}he {E}uropean {N}ucleotide {A}rchive in 2022}}.
\newblock \bibinfo{journal}{\emph{Nucleic Acids Res}} \bibinfo{volume}{51},
  \bibinfo{number}{D1} (\bibinfo{date}{Jan} \bibinfo{year}{2023}),
  \bibinfo{pages}{D121--D125}.
\newblock


\bibitem[Crespo-Cepeda et~al\mbox{.}(2019)]%
        {cepeda2019challenges}
\bibfield{author}{\bibinfo{person}{Rodrigo Crespo-Cepeda},
  \bibinfo{person}{Giuseppe Agapito}, \bibinfo{person}{Jose Vazquez-Poletti},
  {and} \bibinfo{person}{Mario Cannataro}.} \bibinfo{year}{2019}\natexlab{}.
\newblock \showarticletitle{Challenges and Opportunities of Amazon Serverless
  Lambda Services in Bioinformatics}. \bibinfo{pages}{663--668}.
\newblock
\showISBNx{978-1-4503-6666-3}
\urldef\tempurl%
\url{https://doi.org/10.1145/3307339.3343462}
\showDOI{\tempurl}


\bibitem[Di~Tommaso et~al\mbox{.}(2017)]%
        {di2017nextflow}
\bibfield{author}{\bibinfo{person}{Paolo Di~Tommaso}, \bibinfo{person}{Maria
  Chatzou}, \bibinfo{person}{Evan~W Floden}, \bibinfo{person}{Pablo~Prieto
  Barja}, \bibinfo{person}{Emilio Palumbo}, {and} \bibinfo{person}{Cedric
  Notredame}.} \bibinfo{year}{2017}\natexlab{}.
\newblock \showarticletitle{Nextflow enables reproducible computational
  workflows}.
\newblock \bibinfo{journal}{\emph{Nature biotechnology}} \bibinfo{volume}{35},
  \bibinfo{number}{4} (\bibinfo{year}{2017}), \bibinfo{pages}{316--319}.
\newblock


\bibitem[Ferretti et~al\mbox{.}(2019)]%
        {sinple}
\bibfield{author}{\bibinfo{person}{L. Ferretti}, \bibinfo{person}{C.
  Tennakoon}, \bibinfo{person}{A. Silesian}, \bibinfo{person}{G. Freimanis},
  {and} \bibinfo{person}{P. Ribeca}.} \bibinfo{year}{2019}\natexlab{}.
\newblock \showarticletitle{SiNPle: Fast and Sensitive Variant Calling for Deep
  Sequencing Data}.
\newblock \bibinfo{journal}{\emph{Genes}} (\bibinfo{year}{2019}).
\newblock


\bibitem[Hellerstein et~al\mbox{.}(2018)]%
        {hellerstein2018serverless}
\bibfield{author}{\bibinfo{person}{Joseph~M. Hellerstein},
  \bibinfo{person}{Jose Faleiro}, \bibinfo{person}{Joseph~E. Gonzalez},
  \bibinfo{person}{Johann Schleier-Smith}, \bibinfo{person}{Vikram Sreekanti},
  \bibinfo{person}{Alexey Tumanov}, {and} \bibinfo{person}{Chenggang Wu}.}
  \bibinfo{year}{2018}\natexlab{}.
\newblock \bibinfo{title}{Serverless Computing: One Step Forward, Two Steps
  Back}.
\newblock
\newblock
\showeprint[arxiv]{1812.03651}~[cs.DC]


\bibitem[Hung et~al\mbox{.}(2020)]%
        {hong2020accessible}
\bibfield{author}{\bibinfo{person}{Ling-Hong Hung}, \bibinfo{person}{Xingzhi
  Niu}, \bibinfo{person}{Wes Lloyd}, {and} \bibinfo{person}{Ka~Yee Yeung}.}
  \bibinfo{year}{2020}\natexlab{}.
\newblock \showarticletitle{Accessible and interactive RNA sequencing analysis
  using serverless computing}.
\newblock \bibinfo{journal}{\emph{bioRxiv}} (\bibinfo{year}{2020}).
\newblock
\urldef\tempurl%
\url{https://doi.org/10.1101/576199}
\showDOI{\tempurl}
\showeprint{https://www.biorxiv.org/content/early/2020/10/03/576199.full.pdf}


\bibitem[John et~al\mbox{.}(2021)]%
        {aji2021evaluation}
\bibfield{author}{\bibinfo{person}{Aji John}, \bibinfo{person}{Kathleen
  Muenzen}, {and} \bibinfo{person}{Kristiina Ausmees}.}
  \bibinfo{year}{2021}\natexlab{}.
\newblock \showarticletitle{Evaluation of serverless computing for scalable
  execution of a joint variant calling workflow}.
\newblock \bibinfo{journal}{\emph{PLOS ONE}} \bibinfo{volume}{16},
  \bibinfo{number}{7} (\bibinfo{date}{07} \bibinfo{year}{2021}),
  \bibinfo{pages}{1--12}.
\newblock
\urldef\tempurl%
\url{https://doi.org/10.1371/journal.pone.0254363}
\showDOI{\tempurl}


\bibitem[Jonas et~al\mbox{.}(2017)]%
        {jonas2017occupy}
\bibfield{author}{\bibinfo{person}{Eric Jonas}, \bibinfo{person}{Qifan Pu},
  \bibinfo{person}{Shivaram Venkataraman}, \bibinfo{person}{Ion Stoica}, {and}
  \bibinfo{person}{Benjamin Recht}.} \bibinfo{year}{2017}\natexlab{}.
\newblock \showarticletitle{Occupy the Cloud: Distributed Computing for the
  99\%}. In \bibinfo{booktitle}{\emph{Proceedings of the 2017 Symposium on
  Cloud Computing}} (Santa Clara, California) \emph{(\bibinfo{series}{SoCC
  '17})}. \bibinfo{publisher}{Association for Computing Machinery},
  \bibinfo{address}{New York, NY, USA}, \bibinfo{pages}{445–451}.
\newblock
\showISBNx{9781450350280}
\urldef\tempurl%
\url{https://doi.org/10.1145/3127479.3128601}
\showDOI{\tempurl}


\bibitem[Lee et~al\mbox{.}(2019)]%
        {lee2019dnavisualization}
\bibfield{author}{\bibinfo{person}{Benjamin~D Lee}, \bibinfo{person}{Michael~A
  Timony}, {and} \bibinfo{person}{Pablo Ruiz}.}
  \bibinfo{year}{2019}\natexlab{}.
\newblock \showarticletitle{{DNAvisualization.org}: a serverless web tool for
  {DNA} sequence visualization}.
\newblock \bibinfo{journal}{\emph{Nucleic Acids Res}} \bibinfo{volume}{47},
  \bibinfo{number}{W1} (\bibinfo{date}{July} \bibinfo{year}{2019}),
  \bibinfo{pages}{W20--W25}.
\newblock


\bibitem[Marco-Sola et~al\mbox{.}(2012)]%
        {sola2012gem}
\bibfield{author}{\bibinfo{person}{Santiago Marco-Sola},
  \bibinfo{person}{Michael Sammeth}, \bibinfo{person}{Roderic Guig{\'o}}, {and}
  \bibinfo{person}{Paolo Ribeca}.} \bibinfo{year}{2012}\natexlab{}.
\newblock \showarticletitle{The GEM mapper: fast, accurate and versatile
  alignment by filtration}.
\newblock \bibinfo{journal}{\emph{Nature methods}} \bibinfo{volume}{9},
  \bibinfo{number}{12} (\bibinfo{year}{2012}), \bibinfo{pages}{1185--1188}.
\newblock


\bibitem[Niu et~al\mbox{.}(2019)]%
        {niu2019leveraging}
\bibfield{author}{\bibinfo{person}{Xingzhi Niu}, \bibinfo{person}{Dimitar
  Kumanov}, \bibinfo{person}{Ling-Hong Hung}, \bibinfo{person}{Wesley Lloyd},
  {and} \bibinfo{person}{Ka~Yee Yeung}.} \bibinfo{year}{2019}\natexlab{}.
\newblock \showarticletitle{Leveraging Serverless Computing to Improve
  Performance for Sequence Comparison}. \bibinfo{pages}{683--687}.
\newblock
\showISBNx{978-1-4503-6666-3}
\urldef\tempurl%
\url{https://doi.org/10.1145/3307339.3343465}
\showDOI{\tempurl}


\bibitem[Pu et~al\mbox{.}(2019)]%
        {pu2019shuffling}
\bibfield{author}{\bibinfo{person}{Qifan Pu}, \bibinfo{person}{Shivaram
  Venkataraman}, {and} \bibinfo{person}{Ion Stoica}.}
  \bibinfo{year}{2019}\natexlab{}.
\newblock \showarticletitle{Shuffling, Fast and Slow: Scalable Analytics on
  Serverless Infrastructure}. In \bibinfo{booktitle}{\emph{16th USENIX
  Symposium on Networked Systems Design and Implementation (NSDI 19)}}.
  \bibinfo{publisher}{USENIX Association}, \bibinfo{address}{Boston, MA},
  \bibinfo{pages}{193--206}.
\newblock
\showISBNx{978-1-931971-49-2}
\urldef\tempurl%
\url{https://www.usenix.org/conference/nsdi19/presentation/pu}
\showURL{%
\tempurl}


\bibitem[Rodgers(1985)]%
        {rodgers1985amdhal}
\bibfield{author}{\bibinfo{person}{David~P. Rodgers}.}
  \bibinfo{year}{1985}\natexlab{}.
\newblock \showarticletitle{Improvements in Multiprocessor System Design}.
\newblock \bibinfo{journal}{\emph{SIGARCH Comput. Archit. News}}
  \bibinfo{volume}{13}, \bibinfo{number}{3} (\bibinfo{date}{jun}
  \bibinfo{year}{1985}), \bibinfo{pages}{225–231}.
\newblock
\showISSN{0163-5964}
\urldef\tempurl%
\url{https://doi.org/10.1145/327070.327215}
\showDOI{\tempurl}


\bibitem[Sampe et~al\mbox{.}(2021)]%
        {sampe2021lithops}
\bibfield{author}{\bibinfo{person}{Josep Sampe}, \bibinfo{person}{Pedro
  Garcia-Lopez}, \bibinfo{person}{Marc Sanchez-Artigas}, \bibinfo{person}{Gil
  Vernik}, \bibinfo{person}{Pol Roca-Llaberia}, {and} \bibinfo{person}{Aitor
  Arjona}.} \bibinfo{year}{2021}\natexlab{}.
\newblock \showarticletitle{Toward Multicloud Access Transparency in Serverless
  Computing}.
\newblock \bibinfo{journal}{\emph{IEEE Software}} \bibinfo{volume}{38},
  \bibinfo{number}{1} (\bibinfo{year}{2021}), \bibinfo{pages}{68--74}.
\newblock
\urldef\tempurl%
\url{https://doi.org/10.1109/MS.2020.3029994}
\showDOI{\tempurl}


\bibitem[Shankar et~al\mbox{.}(2018)]%
        {shankar2018numpywren}
\bibfield{author}{\bibinfo{person}{Vaishaal Shankar}, \bibinfo{person}{Karl
  Krauth}, \bibinfo{person}{Qifan Pu}, \bibinfo{person}{Eric Jonas},
  \bibinfo{person}{Shivaram Venkataraman}, \bibinfo{person}{Ion Stoica},
  \bibinfo{person}{Benjamin Recht}, {and} \bibinfo{person}{Jonathan
  Ragan-Kelley}.} \bibinfo{year}{2018}\natexlab{}.
\newblock \bibinfo{title}{numpywren: serverless linear algebra}.
\newblock
\newblock
\showeprint[arxiv]{1810.09679}~[cs.DC]


\bibitem[Spillner et~al\mbox{.}(2018)]%
        {spillner2018faaster}
\bibfield{author}{\bibinfo{person}{Josef Spillner}, \bibinfo{person}{Cristian
  Mateos}, {and} \bibinfo{person}{David~A. Monge}.}
  \bibinfo{year}{2018}\natexlab{}.
\newblock \showarticletitle{FaaSter, Better, Cheaper: The Prospect of
  Serverless Scientific Computing and HPC}. In \bibinfo{booktitle}{\emph{High
  Performance Computing}}, \bibfield{editor}{\bibinfo{person}{Esteban Mocskos}
  {and} \bibinfo{person}{Sergio Nesmachnow}} (Eds.).
  \bibinfo{publisher}{Springer International Publishing},
  \bibinfo{address}{Cham}, \bibinfo{pages}{154--168}.
\newblock
\showISBNx{978-3-319-73353-1}


\end{thebibliography}
